\documentclass[10pt,conference]{IEEEtran}
\IEEEoverridecommandlockouts

\usepackage[T1]{fontenc}
\usepackage{lmodern}
\usepackage[utf8]{inputenc}

\usepackage{tabularx}
\usepackage{comment}
\usepackage{amsmath,amssymb,amsfonts}
\usepackage{scalerel}
\usepackage{tikz}\usetikzlibrary{positioning}
\usetikzlibrary{svg.path}
\usepackage{ifthen}
\usepackage{cite}
\usepackage{hyperref}

\definecolor{orcidlogocol}{HTML}{A6CE39}
\tikzset{
	orcidlogo/.pic={
		\fill[orcidlogocol] svg{M256,128c0,70.7-57.3,128-128,128C57.3,256,0,198.7,0,128C0,57.3,57.3,0,128,0C198.7,0,256,57.3,256,128z};
		\fill[white] svg{M86.3,186.2H70.9V79.1h15.4v48.4V186.2z}
		svg{M108.9,79.1h41.6c39.6,0,57,28.3,57,53.6c0,27.5-21.5,53.6-56.8,53.6h-41.8V79.1z M124.3,172.4h24.5c34.9,0,42.9-26.5,42.9-39.7c0-21.5-13.7-39.7-43.7-39.7h-23.7V172.4z}
		svg{M88.7,56.8c0,5.5-4.5,10.1-10.1,10.1c-5.6,0-10.1-4.6-10.1-10.1c0-5.6,4.5-10.1,10.1-10.1C84.2,46.7,88.7,51.3,88.7,56.8z};
	};
}

\newcommand\orcidicon[1]{\href{https://orcid.org/#1}{\mbox{\scalerel*{
			\begin{tikzpicture}[yscale=-1,transform shape]
				\pic{orcidlogo};
			\end{tikzpicture}
		}{|}}}}

\begin{document}
	
	\title{How Do Code Smells Affect Skill Growth in Scratch Novice Programmers?}
	
	\author{
		
		\IEEEauthorblockN{Ricardo Hidalgo Aragón}
		\IEEEauthorblockA{
			\textit{Universidad Rey Juan Carlos}\\
			Madrid, Spain \\
			r.hidalgoa.2024@alumnos.urjc.es}
		\and
		\IEEEauthorblockN{Jesús M. González-Barahona}
		\IEEEauthorblockA{
			\textit{Universidad Rey Juan Carlos}\\
			Madrid, Spain \\
			jesus.gonzalez.barahona@urjc.es}
		\and
		\IEEEauthorblockN{Gregorio Robles}
		\IEEEauthorblockA{
			\textit{Universidad Rey Juan Carlos}\\
			Madrid, Spain \\
			grex@gsyc.urjc.es}
		
	}
	
	\maketitle
	
	\begin{abstract}
		\noindent\textit{Context.} Code smells, which are recurring anomalies in design or style, have been extensively researched in professional code. However, their significance in block-based projects created by novices is still largely unknown. Block‑based environments such as Scratch offer a unique, data‑rich setting to examine how emergent design problems intersect with the cultivation of computational‑thinking (CT) skills.
		\noindent\textit{Objective.} This research explores the connection between CT proficiency and design-level code smells—issues that may hinder software maintenance and evolution—in programs created by Scratch developers. We seek to identify \emph{which} CT dimensions align most strongly with \emph{which} code smells and whether task context moderates those associations.
		\noindent\textit{Method.} A random sample of $\sim$2 million public Scratch projects is mined. Using open-source linters, we extract nine CT scores and 40 code smell indicators from these projects.  After rigorous pre‑processing, we apply descriptive analytics, robust correlation tests, stratified cross-validation, and exploratory machine-learning models; qualitative spot-checks contextualize quantitative patterns.
		\noindent\textit{Impact.} The study will deliver the first large-scale, fine-grained map linking specific CT competencies to concrete design flaws and antipatterns.  Results are poised to (i) inform evidence-based curricula and automated feedback systems, (ii) provide effect-size benchmarks for future educational interventions, and (iii) supply an open, pseudonymized dataset and reproducible analysis pipeline for the research community.  By clarifying how programming habits influence early skill acquisition, the work advances both computing-education theory and practical tooling for sustainable software maintenance and evolution.
	\end{abstract}

	
	
	\section{Introduction}
	\label{sec:intro}
	
	Creating software that is comprehensible, maintainable, and evolvable is a cornerstone of modern software engineering.  Although functional bugs attract immediate attention, \emph{code smells}—recurring structures that violate accepted design or style principles—often foreshadow deeper quality and sustainability problems that impact software maintenance and evolution~\cite{olbrich2009evolution}.  Over the past two decades, researchers have cataloged code smells (CS) in a wide variety of professional code bases, proposed automated detectors, and examined their impact on long‑term maintenance effort.  Yet one community remains largely unexplored: \emph{novice programmers} who are still acquiring fundamental problem‑solving and design skills.
	
	Block‑based environments such as Scratch\footnote{\url{https://scratch.mit.edu/}} lower the entry barrier to programming and are now ubiquitous in K‑12 and introductory university courses~\cite{batni2025current}.  These platforms generate vast public repositories of student work and are accompanied by automated analysis tools—including DrScratch\footnote{\url{https://drscratch.org/}}~\cite{vargas2019bad} and LitterBox\footnote{\url{https://scratch.fim.uni-passau.de/litterbox/}}~\cite{LitterBox_linter}—that extract both computational‑thinking (CT) metrics and CS inventories from project snapshots.  Such data create a timely opportunity to investigate the relationship between emergent design problems and the development of programming competence.
	
	\textbf{Who counts as a novice?} In line with Resnick~\cite{resnick2009scratch} and Dasgupta~\cite{dasgupta2016remixing}, we define a novice Scratch programmer as a creator with little or no prior text-based coding experience (K–12 or first-year undergraduate). Age metadata are unavailable; hence we use Scratch’s intended educational positioning as a proxy for user experience level. While we acknowledge that Scratch is used by both novice programmers and experienced developers--serving as an initiation platform for the former~\cite{resnick2009scratch}, and as a means to explore computational ideas or build advanced creative prototypes for the latter~\cite{dasgupta2016remixing}--this predominant bipolarity justifies RQ1 and RQ2 for beginners and RQ3 for more experienced users.	
	
	\textbf{Maintenance focus.} Prior studies suggest maintenance topics receive comparatively less attention in many introductory courses \cite{gallagher_teaching_2019}.
	
	In remix-based environments such as Scratch, even seemingly benign design flaws---such as cluttered block arrangements, duplicated logic, or unclear naming---can escalate into serious liabilities when propagated through numerous downstream projects. These \textit{liabilities} manifest as increased cognitive load, maintenance complexity, and decreased reusability for remixers. Empirical studies show that when poorly structured code is inherited, novices often replicate or misunderstand flawed patterns, amplifying long-term maintenance costs~\cite{dasgupta2016remixing}. Thus, what appears as a minor issue in an isolated script may result in systemic degradation across a collaborative ecosystem.
	
	\textbf{Research gap.} Prior empirical studies have shown that certain CS surface even in simple Scratch projects \cite{hermans2016smells}, and controlled experiments hint that their presence may hinder code comprehension among beginners \cite{hermans2016code}.  However, we still lack large‑scale evidence clarifying \emph{how} and \emph{which} CS correlate with specific CT dimensions or broader indicators of programming skill, and whether those CS propagate \emph{when} projects are remixed—i.e.: extending \emph{coding challenges} is an instance of when genuine maintenance and evolution occur in the wild.
	Without that evidence, educators and tool builders cannot target feedback where it is most pedagogically and evolutionary valuable.  
	
	\textbf{Objective and scope.} This registered study investigates the associations between (i) nine CT dimensions operationalised by DrScratch and (ii) a catalogue of forty CS detectable by linters. Drawing on a random sample of 2 million public Scratch projects, we will (1) measure the global correlation between overall CT proficiency and overall CS density, (2) identify CT dimensions that are especially predictive of particular CS families, and (3) analyse whether popular block‑programming ``coding challenges'' (i.e.: recreations of \emph{Arkanoid} or \emph{Pac‑Man}, etc.) exhibit distinctive CS profiles.
	
	\textbf{Contributions.} The work advances the literature in three ways:
	
	\begin{enumerate}
		\item \emph{Empirical map}: a comprehensive, statistically‑powered overview of how individual CT competencies align with concrete design antipatterns in novice code.
		\item \emph{Pedagogical insight}: evidence‑based recommendations that enable instructors to tailor exercises, formative assessment, and automated feedback to the smell patterns most indicative of specific learning hurdles.
		\item \emph{Methodological asset}: an openly‑described pipeline that combines large‑scale repository mining with multi‑tool quality assessment, paving the way for future longitudinal or intervention studies on programming‑skill acquisition.
	\end{enumerate}

	By clarifying the links between programming habits and emergent quality problems at scale, this study aims to inform curricula that foster not only functional but also sustainable coding practices from a learner’s very first lines of code.
	
	
	\section{Research Questions}
	\label{sec:obj}
	
	We decompose the objective into one \emph{root} question and three concrete sub‑questions.  
	Each sub‑question is accompanied by its rationale, the planned analytic strategy (summarised here and detailed in Section~\ref{sec:meth}), the anticipated empirical signal that would address it, and the null and alternative hypothesis.
	
	\noindent\textbf{Root Question (RQ‑0).} \textit{How strongly are overall CT proficiency and overall CS density correlated in Scratch projects?}
	
	To address the root question rigorously, we structure the investigation around three analytical axes that progressively refine the granularity of the analysis. First, we examine global patterns between overall CT proficiency and aggregate CS prevalence—what we term the \emph{Aggregate Association} level. Second, we zoom into more fine‑grained pairings between specific CT dimensions and individual or grouped CS, to explore \emph{Dimension–CS Specificity}. Finally, we recognize that the nature of the programming task may itself influence CS incidence. Hence, we examine whether known ``coding challenges'' introduce confounding patterns or moderate the strength of the observed associations—an angle we refer to as \emph{Task Context Moderation}. Table~\ref{tab:analysis_axes} summarizes these three perspectives, the corresponding unit of analysis, and the interpretive focus.
	
	\begin{table}[htbp]
		\caption{Analytical Axes for Addressing the Root Research Question}
		\label{tab:analysis_axes}
		\centering
		\begin{tabularx}{\columnwidth}{|X|X|X|}
			\hline
			\textbf{Analytical Axis} & \textbf{Unit of Analysis} & \textbf{Interpretive Focus} \\
			\hline
			Aggregate Association & Project‑level summary (CT score vs.\ CS density) & Do more skilled programmers produce less CS overall? \\
			\hline
			Dimension–CS Specificity & CT dimensions and CS categories & Which CT competencies mitigate which kinds of design flaws? \\
			\hline
			Task Context Moderation & Project type (i.e., challenge vs.\ non‑challenge) & Are CS patterns explained by task difficulty rather than user skill? \\
			\hline
		\end{tabularx}
	\end{table}
	
	Considering the analytical axes described above, our proposed RQs are as follows:
	
	\noindent\textbf{RQ‑1 -- Does higher overall CT proficiency predict a lower density of code smells?} \textit{[Aggregate Association]}
	\begin{itemize}
		\item{Motivation:} If CT skills support better design choices, projects with stronger CT scores should exhibit fewer CS.  
		\item{Operational plan:} Compute (a) the sum of the nine CT scores and (b) the total CS count normalised by block count; test the relationship with Spearman’s $\rho$ and robust regression. Boosting models could be used to define most relevant features. 
		\item{Expected evidence:} A negative—but not necessarily linear—association indicating that improved CT is linked to cleaner design.
		\item{H$_0$:} $\rho \ge 0$ (no association or non-negative association).
		\item{H$_1$:} $\rho < 0$ (higher CT is associated with lower CS density).	
	\end{itemize}
	
	\noindent\textbf{RQ‑2 -- Which CT dimensions are most predictive of particular code smell families?}  \textit{[Dimension–Code Smell Specificity]}
	\begin{itemize}
		\item{Motivation:} Certain cognitive skills~\cite{keuning2023systematic} may guard against specific antipatterns~\cite{paiva2025improving}. Identifying these links enables targeted pedagogy.  
		\item{Operational plan:} For each CT dimension, fit generalized linear models and apply mutual‑information ranking against grouped CS; corroborate with hierarchical clustering. To enhance model explainability, apply LIME/SHAP~\cite{islam2024enhancing} to identify and visualize the local and global contribution of CT dimensions to the prediction of each CS family.
		\item{Expected evidence:} A ranked matrix showing, for instance, that low \emph{Abstraction} scores coincide with cloning‑related CS, whereas weak \emph{Flow Control} relates to busy‑waiting. Complementary LIME/SHAP techniques will provide both global feature impact rankings and local explanations for individual projects, confirming which CT dimensions most strongly influence each CS and under what conditions.
		\item{H$_0$:} For every CT dimension $d$ and CS family $s$, the mutual information $I(d;s)=0$ \emph{and} the GLM coefficient $\beta_{d,s}=0$; that is, no CT dimension contributes predictive information about the presence or intensity of any CS family.
		\item{H$_1$:} There exists at least one pair $(d,s)$ such that $I(d;s)>0$ \emph{or} $\beta_{d,s}\neq 0$; that is, at least one CT dimension provides statistically significant predictive power for at least one CS family.
		\item Here, $I(d;s)$ denotes the mutual information between CT dimension $d$ and the occurrence/intensity of CS family $s$ (assessed via $\chi^{2}$ or permutation tests), while $\beta_{d,s}$ is the coefficient of $d$ in the generalized linear model for $s$ (evaluated with Wald or $z$ tests).
	\end{itemize}
	
	\noindent\textbf{RQ‑3 -- Does project context (generic projects vs.\ defined ``coding challenges'') moderate the CT–CS relationship?}  \textit{[Task Context Moderation]}
	\begin{itemize}
		\item{Motivation:} Some Code Smells may arise from the intrinsic complexity of certain challenges rather than from misunderstandings. Distinguishing context‑driven code smells prevents misinterpreting skill level.  
		\item{Operational plan:} Automatically label projects that match prototypical challenge patterns, most of which should come from \texttt{remix} actions; run the RQ‑1 and RQ‑2 analyses within and across context strata; test interaction terms.  
		\item{Expected evidence:} Either (i) similar associations across contexts—implying CS mostly reflect skill—or (ii) significant interaction effects that flag context‑specific CS (i.e.: polling loops in fast‑paced games).
		\item{H$_0$:} The CT–CS relationship is \emph{not} moderated by
		project context; formally, the interaction coefficient
		$\gamma_{\text{CT}\times\text{Context}} = 0$
		(or, equivalently, $\rho_{\text{generic}} = \rho_{\text{challenge}}$ for every CS family).
		\item{H$_1$:} At least one moderation effect exists; that is,
		$\gamma_{\text{CT}\times\text{Context}} \neq 0$
		(or, for at least one CS family,
		$\rho_{\text{generic}} \neq \rho_{\text{challenge}}$),
		indicating that project context alters the strength or direction
		of the CT–CS association.
	\end{itemize}
	
	\noindent\emph{Scope note.}  
	All analyses are cross‑sectional; we use multiple snapshots rather than longitudinal traces, so \emph{development} refers to \emph{relative proficiency} inferred from completed projects, not to time‑series skill gain. No causal claims will be made.
	
	\section{Background}
	\label{sec:background}
	
	This section outlines the theoretical background of our study, and identifies the gap this study aims to fill.
	
	\subsection{Block‑Based Programming and the Rationale for Using Scratch}
	
	Block‑based environments employ a “puzzle‑piece’’ visual syntax that constrains how instructions can be combined, reducing incidental syntactic errors while foregrounding sequencing and logic \cite{weintrop2019block}. Among them, Scratch stands out as the most suitable platform for large-scale repository mining due to its extensive public corpus of over 164 million projects\footnote{\url{https://scratch.mit.edu/statistics/}}, support for automated data access via REST API, and the availability of mature, open-source tools for .sb3 file analysis and design-CS detection. In contrast, alternative environments like Snap!\footnote{\url{https://snap.berkeley.edu/}} and Blockly\footnote{\url{https://blockly.games/}} lack the necessary scale, openness, and tooling. Recent curricular analyses~\cite{stewart2023analyzing} confirm Scratch’s global dominance in introductory computer science education.
	
	\subsection{Computational Thinking: Concepts and Automated Assessment}
	
	Wing’s conceptualization of computational thinking as a transferable, problem-solving disposition has catalyzed widespread educational interest and research, bridging computer science with broader cognitive skill development~\cite{lodi_computational_2021}.
	
	Contemporary psychometric efforts tend to operationalize Computational Thinking through specific dimensions such as abstraction, parallelism, and control flow, which have demonstrated psychometric validity in various studies~\cite{roman2019assessment, zhang2023cttest}.       
	
	In the Scratch ecosystem, DrScratch provides a nine‑dimension rubric scored automatically from project code blocks: \emph{Abstraction, Parallelism, Logic, Synchronisation, Flow Control, User Interactivity, Data Representation, Math Operators,} and \emph{Motion Operators}.  Validation studies report moderate agreement with expert ratings \cite{alves2019approaches} and growing adoption in empirical CT work.  Automated CT scoring enables scalable, reproducible analysis without observer bias—an essential prerequisite for correlational studies involving millions of artifacts.
	
	\subsection{Code Smells in Novice and Block‑Based Software}
	
	Code smells are structural or stylistic patterns that hinder code comprehension and maintenance \cite{fowler2018refactoring}. In block languages, early work identified code smells like duplicated scripts and unused blocks \cite{hermans2016smells}. Studies show some CS affect novice understanding \cite{hermans2016code}, and tools like DrScratch and LitterBox now analyze millions of projects with classroom‑ready feedback \cite{moreno2014automatic, LitterBox_linter}. Recent research explores smell clustering and machine learning prediction \cite{santana2024unraveling}, though educational impact remains underexplored.

	\subsection{CS as Predictors of Maintenance and Evolution Effort}
	\label{sec:maintenance}
	
	Classic maintenance research demonstrates that persistent CS increases change-proneness, bug‑fix cost, and developer churn in professional systems~\cite{fowler2018refactoring,olbrich2009evolution}
	
	\subsection{Synthesis and Research Gap}
	
	Two observations motivate our work. First, CT metrics and smell detectors now coexist, yet they have rarely been examined in concert. Second, while CS are hypothesised to reflect deeper misconceptions, no large-scale evidence clarifies \textit{which} CT competencies (if any) are most strongly associated \textit{with} particular CS, nor \textit{how} task context moderates that association. The present study addresses this gap by statistically mapping CT-smell relationships across a random sample of $\sim$2 million Scratch projects, thereby bridging educational-computing and software-quality research.
	
	
	\section{Study Design and Dataset}
	\label{sec:meth}
	
	\subsection{Overview}
	
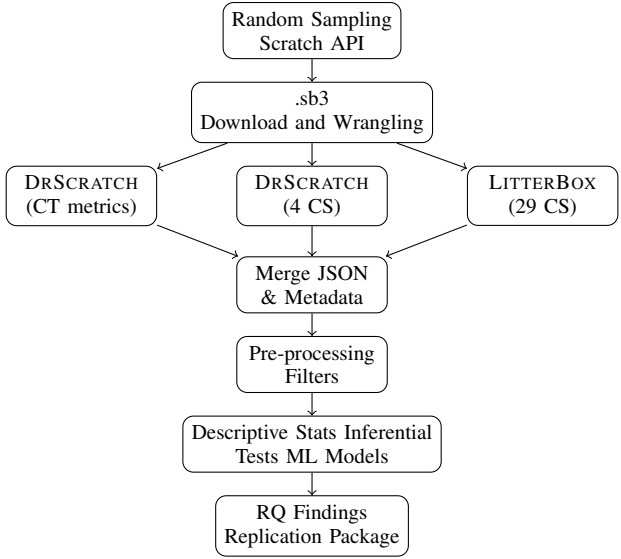
\begin{figure}[t]
		\centering
		\begin{tikzpicture}[
			node distance = 3mm and 2mm,
			box/.style = {draw, rounded corners, align=center,
				minimum height=4mm, minimum width=20mm,
				font=\footnotesize}]
			\node[box] (sample)  {Random Sampling\\Scratch API};
			\node[box, below=of sample] (download) {.sb3\\Download and Wrangling};
			\node[box, below left=of download, xshift=-2.5mm] (dr)
			{\textsc{DrScratch}\\(CT metrics)};
			\node[box, below=of download] (lb)
			{\textsc{DrScratch}\\(4 CS)};
			\node[box, below right=of download, xshift=2.5mm] (qh)
			{\textsc{LitterBox}\\(29 CS)};
			\node[box, below=4mm of lb] (merge)
			{Merge JSON\\\& Metadata};
			\node[box, below=of merge] (clean)
			{Pre‑processing\\Filters};
			\node[box, below=of clean] (analysis)
			{Descriptive Stats Inferential\\Tests ML Models};
			\node[box, below=of analysis] (outputs)
			{RQ Findings\\Replication Package};
			\draw[->] (sample)  -- (download);
			\draw[->] (download) -- (dr);
			\draw[->] (download) -- (lb);
			\draw[->] (download) -- (qh);
			\draw[->] (dr)       -- (merge);
			\draw[->] (lb)       -- (merge);
			\draw[->] (qh)       -- (merge);
			\draw[->] (merge)    -- (clean);
			\draw[->] (clean)    -- (analysis);
			\draw[->] (analysis) -- (outputs);
		\end{tikzpicture}
		\caption{End‑to‑end data‑collection and analysis pipeline used in this study.}
		\label{fig:pipeline}
	\end{figure}

	To answer the research questions articulated in Section~\ref{sec:obj}, we adopt a \emph{cross-sectional, repository-mining design} informed by prior standards~\cite{nagappan2006mining}, and a \emph{data-intensive approach} following Ramasamy et al.~\cite{ramasamy_workflow_2022}.
	A large, randomly sampled corpus of public Scratch projects is mined, enriched with CT metrics and CS, then analyzed through a combination of descriptive statistics, correlational tests, and exploratory machine-learning models. A schematic of the end-to-end pipeline is given in Figure~\ref{fig:pipeline}.
	
	\subsection{Data Source and Sampling Strategy}
	\label{subsec:data_source}
	
	\begin{itemize}
		\item \textbf{Population.} All public projects hosted on the official Scratch portal.  
		\item \textbf{Sampling frame.} 
		\begin{itemize}
			\item We will query the Scratch REST API and retrieve project identifiers using uniform random offsets in the search index to avoid chronological or popularity bias.  
			\item We discard empty, duplicate and byte-identical projects; remixes are
			kept and flagged for RQ 3.
			\item Random API offsets guarantee an equal inclusion probability for each of the $\approx$164 M public projects.
		\end{itemize}
		\item \textbf{Target sample size.} $\sim$2 million distinct projects—a scale shown to provide stable effect-size estimates in pilot power analyses (Section~\ref{subsec:power}). This size balances statistical power with storage ($\approx$500 GB) and processing constraints on our cluster.  
		\item \textbf{Versioning.} For each project we capture the latest revision available at crawl time; longitudinal trajectories are out of scope for the present registered report but will be logged (remixed's info) for future work.
	\end{itemize}
	
	\subsection{Automated Feature Extraction}
	\label{subsec:feature_extraction}
	
	We perform static analysis using three open-source linters, executed in batch mode, if available via Docker containers:
	
	\begin{itemize}
		\item \textbf{DrScratch}: Computes 9 CT dimensions (extended rubric, 0--4 scale) and detects 4 structural code smells.
		\item \textbf{LitterBox}: Detects 29 extra block-specific CS.
	\end{itemize}
	
	The complete list of analyzed CS is presented in Table~\ref{tab:tab_bad_smells}.
	
	\begin{table}[htbp]
		\caption{Code Smells Detectable by the Linters}
		\begin{center}
			\begin{tabularx}{\columnwidth}{|X|X|}
				\hline
				\multicolumn{2}{|c|}{\textbf{Detected Code Smells}} \\
				\hline
				Duplicated Scripts\textsuperscript{(1)} & Unnecessary Middle Man\textsuperscript{(2)} \\
				\hline
				Sprite naming\textsuperscript{(1)} & Multiple Attribute Modifications\textsuperscript{(2)} \\
				\hline
				Backdrop naming\textsuperscript{(1)} & Nested Loops\textsuperscript{(2)} \\
				\hline
				Dead code\textsuperscript{(1)} & Same Variable in Different Sprite\textsuperscript{(2)} \\
				\hline
				Ambiguous Parameter Name Unused\textsuperscript{(2)} & Sequence of Repeated Actions\textsuperscript{(2)} \\
				\hline
				Busy Waiting\textsuperscript{(2)} & Unnecessary If\textsuperscript{(2)} \\
				\hline
				Duplicated Code\textsuperscript{(2)} & Unnecessary If after Repeat-Until\textsuperscript{(2)} \\
				\hline
				Identical Statements\textsuperscript{(2)} & Unnecessary Loop\textsuperscript{(2)} \\
				\hline
				Similar Code\textsuperscript{(2)} & Unnecessary Message\textsuperscript{(2)} \\
				\hline
				Code Lying Around\textsuperscript{(2)} & Unnecessary Time Block\textsuperscript{(2)} \\
				\hline
				Duplicated If-Condition\textsuperscript{(2)} & Unused Custom Block\textsuperscript{(2)} \\
				\hline
				Unused Parameter\textsuperscript{(2)} & Unused Variable\textsuperscript{(2)} \\
				\hline
				Duplicated Sprite\textsuperscript{(2)} & Variable Initialization Race Condition\textsuperscript{(2)} \\
				\hline
				Empty Control Body\textsuperscript{(2)} & Empty Custom Block\textsuperscript{(2)} \\
				\hline
				Empty Project\textsuperscript{(2)} & Empty Script\textsuperscript{(2)} \\
				\hline
				Empty Sprite\textsuperscript{(2)} & Long Script\textsuperscript{(2)} \\
				\hline
				Message Naming\textsuperscript{(2)} & \\
				\hline
			\end{tabularx}
			\label{tab:tab_bad_smells}
			
			\begin{flushleft}
				Analyzed by:
				\textsuperscript{(1)}DrScratch, or \textsuperscript{(2)}LitterBox
			\end{flushleft}
		\end{center}
	\end{table}
	
	Tool outputs (JSON) are merged by a Python script using project ID and enriched with basic metadata.
	
	We enforce data quality by removing empty/duplicate projects, normalizing smell counts by block total, and hashing nicknames.
	
	\subsection{Dataset Schema}
	\label{subsec:schema}
	
	Table~\ref{tab:dataschema} summarizes the final relational dataset containing one row per project and more than 175 derived variables.
	
	\begin{table}[htbp]
		\centering
		\caption{Core Variables in the Unified Study Dataset}
		\label{tab:dataschema}
		\begin{tabular}{|p{2.2cm}|p{5.4cm}|}
			\hline
			\textbf{Variable family} & \textbf{Fields (examples)} \\
			\hline
			Project metadata & \texttt{proj\_id}, hashed \texttt{author\_id}, \texttt{created\_at}, block count, sprite count, project title \\[2pt]
			CT metrics & 9 dimensions (i.e.: \textit{Abstraction\_score}, \textit{FlowCtrl\_score}) from DrScratch \\[2pt]
			Smell metrics & 40 binary or count features (i.e.: \textit{DuplicatedScript}, \textit{BusyWaiting}) aggregated per tool \\[2pt]
			Context labels & Boolean flags for \emph{coding challenges} (Arkanoid-like, Pac-Man-like, etc.) identified via regex patterns and manual validation \\[2pt]
			Derived ratios & CS per 100 blocks, CT sum score, block density per sprite, etc. \\
			\hline
		\end{tabular}
	\end{table}
	
	\subsection{Analysis Pipeline}
	\label{subsec:analysis}
	
	\paragraph*{Descriptive statistics} Block counts, CT scores, and smell densities will be profiled with medians, median absolute deviations (MADs), and kernel–density plots.
	
	\paragraph*{Inferential tests}  
	\begin{itemize}
		\item \textbf{RQ-1 (aggregate association):} Spearman’s $\rho$ between total CT score and smell density; robust linear regression with Huber loss~\cite{huang_robust_2021} to obtain effect size ($\beta$) and 95 \% CIs; and boosting techniques.
		\item \textbf{RQ-2 (dimension–smell specificity):} For every CT dimension $d$ and smell family $s$ we test	$H_{0}\!: I(d;s)=0 \land \beta_{d,s}=0$ versus $H_{1}\!: I(d;s)>0 \lor \beta_{d,s}\neq0$.  
		Concretely, we (i) compute permutation-based mutual information	$I(d;s)$ and (ii) fit generalized linear models—negative-binomial for smell counts and logistic for smell presence—evaluating the coefficient $\beta_{d,s}$ with Wald $z$ tests.  
		All $p$-values are FDR-controlled at $0.05$; SHAP visualize global and local importance.
		\item \textbf{RQ-3 (task-context moderation):} Smell density (and, analogously, individual CS) is modeled with a GLM containing the continuous CT score, a binary	\textit{Context} factor (generic vs.\ challenge), and their interaction term $\gamma_{\\text{CT}\times\\text{Context}}$.  
		The null $H_{0}\!: \gamma = 0$ is tested with Wald $z$ or a likelihood-ratio test, followed by FDR correction across smell families.  
		For robustness we additionally compare context-stratified Spearman correlations using Fisher’s $z$ transformation.
	\end{itemize}
	
	\paragraph*{Exploratory ML} A suite of clustering, regression, and boosting algorithms over the
	combined CT-and-smell feature space (scaled with \texttt{RobustScaler})	will probe latent project archetypes.
	
	\paragraph*{Qualitative mini-case study} We will perform open coding under a grounded-theory lens on 30 projects (10 per extreme quantile).
	
	\subsection{Validation and Statistical Power}
	\label{subsec:power}
	
	A priori power simulation (10,000 draws) indicates that with $N = 1,000,000$ we have $>99\%$ power to detect a small Spearman correlation of $|\rho|=0.05$ at $\alpha=0.001$~\cite{cohen2013statistical}.  All predictive models will employ 5 fold cross-validation stratified by challenge type (RQ3); performance metrics (RMSE, AUROC) will be averaged across folds.  Bootstrapped confidence intervals ($B = 2,000$) will accompany all key statistics~\cite{tibshirani1993introduction}.
	
	\subsection{Ethical and Privacy Considerations}
	
	We use only public artefacts, treating Scratch nicknames as personal data (stored as salted hashes); no raw IDs are published, and analyses are aggregate. Processing relies on \textit{legitimate interest for scientific research} (GDPR Art.~6(1)(f), Art.~89).
	
	\noindent\textbf{Replication package.}  All scripts, container manifests, and aggregate (hashed) datasets will be released under an MIT license upon Stage‑2 acceptance to ensure full reproducibility.
	
	
	\section{Threats to Validity}
	\label{sec:threats}
	
	Consistent with best practice~\cite{ACM_SIGSOFT_Empirical_Standards} in empirical software engineering studies, we discuss potential threats that could compromise the interpretability of our findings and outline the concrete steps we will take to mitigate them.  Table~\ref{tab:threat_matrix} provides a concise overview; the text below elaborates on each category.
	
	\begin{table}[htbp]
		\centering
		\caption{Summary of principal validity threats and planned mitigations.}
		\label{tab:threat_matrix}
		\begin{tabular}{|p{2.5cm}|p{5cm}|}
			\hline
			\textbf{Threat category} & \textbf{Key mitigations} \\ \hline
			Construct validity & Dual operationalization of constructs, qualitative spot checks, multi-tool triangulation \\ \hline
			Internal validity & Size normalized smell metrics, model controls for task complexity, stratified resampling \\ \hline
			External validity & 2-million-project random sample, reporting of population coverage limits, replication script release \\ \hline
			Conclusion (statistical) validity & Power simulation, robust estimators, multiple-comparison correction, 5-fold CV \\ \hline
			Reliability of automated tools & Version pinning, manual audit of subsample, open configuration files \\ \hline
			Ethical / privacy risks & SHA-256 hashing of author IDs, aggregate reporting only \\ \hline
		\end{tabular}
	\end{table}
	
	\subsection{Construct Validity}
	
	To reduce bias and improve validity, we (i) triangulate across tools, (ii) conduct qualitative checks on 385 random projects (95\% confidence), and (iii) account for context-sensitive CS that may reflect creativity rather than poor design.
	
	\subsection{Internal Validity}
	
	To reduce confounding (i.e.: project size, visual complexity, challenge difficulty), we normalize smell counts, include key covariates (block/sprite count, challenge type) in regressions, and cluster by latent features to test robustness within subgroups. 
	
	\subsection{External Validity}
	
	Scratch users are not representative of all novices, and findings may not generalize to text-based or professional contexts. To enhance external validity, we sample $\sim$2M projects uniformly via API, publish dataset descriptors, and release anonymized data and scripts for replication.
	
	\subsection{Conclusion (Statistical) Validity}
	
	To mitigate risks of low power, model bias, and multiple testing, we show via 10{,}000-draw Monte Carlo that \(N=10^6\) yields \(>99\%\) power for \(|\rho|=0.05\) at \(\alpha=0.001\). We use robust estimators, FDR correction (Benjamini--Hochberg), and 5-fold cross-validation with full performance reporting.
	
	\subsection{Reliability of Automated Tools}
	
	Static analyzers may evolve or err. To ensure transparency, we fix tool versions, publish configs and logs.
	
	
	\section{Feasibility Study}
	\label{sec:feas}
	
	The study is feasible within six months using existing infrastructure, minimal funding, and a clear risk plan.
	
	\textbf{Infrastructure:} 720 node-hours compute, 2TB temp storage (final dataset $\sim$1GB), pinned Docker images (Zenodo), public scripts/CI.
	
	\textbf{Resources:} Team: PhD student, postdoc, PI (0.1 FTE). No external funding. Postdoc reviews $\sim$385 projects.
	
	\textbf{Risks:} API capped at 100 req/min (14-day harvest); retries for failures; streaming limits storage to 50\%; full scripting ensures handover.
	
	\textbf{Timeline:} \textit{M0:} Setup; \textit{M1--2:} Harvest; \textit{M3:} Freeze dataset; \textit{M4:} Stats, experiments; \textit{M5:} Case study, drafting; \textit{M6:} Finalize for review.

	\section{Expected Contributions}
	\label{sec:contrib}
	
	This registered report promises a multi-faceted contribution that spans empirical knowledge, pedagogical practice, research infrastructure, and open science.  Table~\ref{tab:contrib_overview} provides a bird’s-eye summary; subsections elaborate the specific advances.
	
	\begin{table}[htbp]
		\centering
		\caption{Overview of the study’s anticipated contributions.}
		\label{tab:contrib_overview}
		\begin{tabular}{|p{2.4cm}|p{5.7cm}|}
			\hline
			\textbf{Dimension} & \textbf{Contribution highlights} \\ \hline
			Empirical evidence & First large-scale, statistically powered map between nine CT dimensions and 40 CS across $\sim$2 M Scratch projects \\ \hline
			Pedagogical impact & Data-driven guidance for instructors, curriculum designers, and tool builders (real-time linter feedback, scaffolded exercises) \\ \hline
			Methodological asset & Re-usable pipeline + rigorous power analysis; template for future RR studies on mined educational data \\ \hline
			Open dataset & Public, pseudonymised CT $+$ smell matrix with documentation and query notebooks \\ \hline
			Theoretical insight & Clarifies how design-level CS—not merely complexity—align with cognitive competencies \\ \hline
			Future research enabler & Identifies smell/CT pairs worth longitudinal tracking; flags challenge-specific confounds to inspire controlled experiments \\ \hline
		\end{tabular}
	\end{table}
	
	\subsection{Empirical Knowledge for the Research Community}
	
	\begin{itemize}
		\item \textbf{Granular correlation atlas.}  By cross-tabulating nine CT scores against 40 smell categories, the study will deliver the most fine-grained correlation matrix to date for block-based code, revealing which competencies (i.e.: \emph{Abstraction}) most strongly guard against which antipatterns (i.e.: duplicated scripts).
		\item \textbf{Context moderation evidence.}  The moderation analysis differentiates smell patterns caused by learner misconceptions from those induced by task complexity (i.e.: fast-paced game loops).
		\item \textbf{Effect-size benchmarks.}  Reporting bootstrapped confidence intervals and variance explained will furnish baseline numbers against which future educational-intervention studies can be evaluated.
	\end{itemize}
	
	\subsection{Pedagogical and Practical Impact}
	
	\begin{itemize}
		\item \textbf{Targeted instructional design.}  Instructors will be able to prioritise CT dimensions that exhibit the strongest linkage to frequent CS, focusing lesson time where the empirical deficit is largest.  
		\item \textbf{Actionable tooling feedback.}  Findings could feed back into formative tools such as LitterBox and DrScratch, allowing real-time warnings to be ranked by pedagogical relevance instead of raw frequency.  
		\item \textbf{Evidence-based challenge curation.}  If specific coding tasks disproportionately trigger certain CS, educators can scaffold or sequence those challenges more effectively. 
		\item \textbf{Maintenance‑literacy insight.} By empirically showing how CS hinder the re‑use and evolution of student projects, the study legitimizes ``maintenance thinking'' in early programming curricula and supplies concrete examples that instructors can use to teach refactoring and sustainable design from day one.
	\end{itemize}
	
	\subsection{Methodological and Infrastructure Contributions}
	
	\begin{itemize}
		\item \textbf{Reusable pipeline.} The software, CI-tested workflow provides a blueprint for mining-and-analysis RRs on educational code repositories.
		\item \textbf{Open, extensible dataset.} A pseudonymised Parquet table containing CT metrics, smell flags, and rich metadata will be deposited under an MIT license, lowering the barrier for secondary studies on learning analytics or smell detection.  
		\item \textbf{Power-analysis template.}  The Monte-Carlo scripts shipped with the replication package exemplify how to justify sample sizes for large-scale mined data.
	\end{itemize}
	
	\subsection{Theoretical Advancement}
	
	By juxtaposing cognitive-skill proxies with higher-level design deficiencies, the study tests the proposition that CS capture misconceptions more directly than traditional complexity metrics.
	
	\subsection{Catalyst for Future Work}
	
	The correlation atlas, contextual annotations, and open code will highlight unanswered questions—i.e.: whether repeated-script CS diminish as learners progress through a semester—and thereby seed longitudinal RRs or intervention trials.  Moreover, the dataset can serve as a training benchmark for machine-learning models that predict novice misconceptions directly from code, aligning with calls for AI-enhanced tutoring systems.
	
	\vspace{0.5em}
	\noindent\textbf{In sum}, this investigation would produce new and practical empirical findings, a scalable research scaffold, and concrete artifacts for the software maintenance and evolution community.
	

	
	\bibliographystyle{IEEEtran}
	\bibliography{bibliog}

\begin{thebibliography}{10}
\providecommand{\url}[1]{#1}
\csname url@samestyle\endcsname
\providecommand{\newblock}{\relax}
\providecommand{\bibinfo}[2]{#2}
\providecommand{\BIBentrySTDinterwordspacing}{\spaceskip=0pt\relax}
\providecommand{\BIBentryALTinterwordstretchfactor}{4}
\providecommand{\BIBentryALTinterwordspacing}{\spaceskip=\fontdimen2\font plus
\BIBentryALTinterwordstretchfactor\fontdimen3\font minus
  \fontdimen4\font\relax}
\providecommand{\BIBforeignlanguage}[2]{{%
\expandafter\ifx\csname l@#1\endcsname\relax
\typeout{** WARNING: IEEEtran.bst: No hyphenation pattern has been}%
\typeout{** loaded for the language `#1'. Using the pattern for}%
\typeout{** the default language instead.}%
\else
\language=\csname l@#1\endcsname
\fi
#2}}
\providecommand{\BIBdecl}{\relax}
\BIBdecl

\bibitem{olbrich2009evolution}
S.~Olbrich, D.~S. Cruzes, V.~Basili, and N.~Zazworka, ``The evolution and
  impact of code smells: A case study of two open source systems,'' in
  \emph{2009 3rd International Symposium on Empirical Software Engineering and
  Measurement}, 2009, pp. 390--400.

\bibitem{batni2025current}
B.~Batni, S.~N. Junaini, J.~Sidi, W.~A. Mustafa, and Z.~I.~A. Ismail, ``Current
  research trends of scratch block based programming for k-12: A systematic
  review,'' \emph{Journal of Advanced Research in Applied Sciences and
  Engineering Technology}, vol.~51, no.~2, pp. 138--152, 2025.

\bibitem{vargas2019bad}
{\'A}.~Vargas-Alba, G.~M. Troiano, Q.~Chen, C.~Harteveld, and G.~Robles, ``Bad
  smells in scratch projects: Prel. analysis,'' in \emph{TACKLE@ EC-TEL}, 2019.

\bibitem{LitterBox_linter}
G.~Fraser, U.~Heuer, N.~Körber, F.~Obermüller, and E.~Wasmeier, ``Litterbox:
  A linter for scratch programs,'' in \emph{2021 IEEE/ACM 43rd International
  Conference on Software Engineering: Software Engineering Ed. and Training
  (ICSE-SEET)}, 2021, pp. 183--188.

\bibitem{resnick2009scratch}
M.~Resnick, J.~Maloney, A.~Monroy-Hern{\'a}ndez, N.~Rusk, E.~Eastmond,
  K.~Brennan, A.~Millner, E.~Rosenbaum, J.~Silver, B.~Silverman \emph{et~al.},
  ``Scratch: programming for all,'' \emph{Communications of the ACM}, vol.~52,
  no.~11, pp. 60--67, 2009.

\bibitem{dasgupta2016remixing}
S.~Dasgupta, W.~Hale, A.~Monroy-Hern{\'a}ndez, and B.~M. Hill, ``Remixing as a
  pathway to computational thinking,'' in \emph{Proceedings of the 19th ACM
  conference on computer-supported cooperative work \& social computing}, 2016,
  pp. 1438--1449.

\bibitem{gallagher_teaching_2019}
K.~Gallagher, M.~Fioravanti, and S.~Kozaitis, ``Teaching {Software}
  {Maintenance},'' in \emph{2019 {IEEE} {International} {Conference} on
  {Software} {Maintenance} and {Evolution} ({ICSME})}, pp. 353--362.

\bibitem{hermans2016smells}
F.~Hermans, K.~T. Stolee, and D.~Hoepelman, ``Smells in block-based programming
  languages,'' in \emph{2016 IEEE Symposium on Visual Languages and
  Human-Centric Computing (VL/HCC)}.\hskip 1em plus 0.5em minus 0.4em\relax
  IEEE, 2016, pp. 68--72.

\bibitem{hermans2016code}
F.~Hermans and E.~Aivaloglou, ``Do code smells hamper novice programming? a
  controlled experiment on scratch programs,'' in \emph{2016 IEEE 24th
  International Conference on Program Comprehension (ICPC)}.\hskip 1em plus
  0.5em minus 0.4em\relax IEEE, 2016, pp. 1--10.

\bibitem{keuning2023systematic}
H.~Keuning, J.~Jeuring, and B.~Heeren, ``A systematic mapping study of code
  quality in education--with complete bibliography,'' \emph{arXiv preprint
  arXiv:2304.13451}, 2023.

\bibitem{paiva2025improving}
\BIBentryALTinterwordspacing
J.~C.~C. Paiva, ``Improving feedback in the automated assessment of programming
  assignments using students' past solutions,'' PhD thesis, University of
  Porto, 2025. [Online]. Available:
  \url{https://repositorio-aberto.up.pt/bitstream/10216/165474/2/711241.pdf}
\BIBentrySTDinterwordspacing

\bibitem{islam2024enhancing}
M.~R. Islam, A.~M. Nitu, M.~A. Marjan, M.~P. Uddin, M.~I. Afjal, and M.~A.~A.
  Mamun, ``Enhancing tertiary students’ programming skills with an expl.
  educational data mining approach,'' \emph{PloS one}, vol.~19, no.~9, p.
  e0307536, 2024.

\bibitem{weintrop2019block}
D.~Weintrop, ``Block-based programming in computer science education,''
  \emph{Communications of the ACM}, vol.~62, no.~8, pp. 22--25, 2019.

\bibitem{stewart2023analyzing}
W.~Stewart and K.~Baek, ``Analyzing computational thinking studies in scratch
  programming: A review of elementary education literature,''
  \emph{International Journal of Computer Science Education in Schools},
  vol.~6, no.~1, pp. 35--58, 2023.

\bibitem{lodi_computational_2021}
M.~Lodi and S.~Martini, ``Computational thinking, between papert and wing,''
  vol.~30, no.~4, pp. 883--908.

\bibitem{roman2019assessment}
M.~Rom{\'a}n-Gonz{\'a}lez, J.~Moreno-Le{\'o}n, and G.~Robles, \emph{Combining
  Assessment Tools for a Comprehensive Evaluation of Computational Thinking
  Interventions}.\hskip 1em plus 0.5em minus 0.4em\relax Springer, 2019, pp.
  79--100.

\bibitem{zhang2023cttest}
S.~Zhang and G.~K.~W. Wong, ``Development and validation of a computational
  thinking test for lower primary school students,'' \emph{Educational
  Technology Research and Development}, 2023.

\bibitem{alves2019approaches}
N.~D.~C. Alves, C.~G. Von~Wangenheim, and J.~C. Hauck, ``Approaches to assess
  computational thinking competences based on code analysis in k-12 education:
  A systematic mapping study,'' \emph{Informatics in Education}, vol.~18,
  no.~1, p.~17, 2019.

\bibitem{fowler2018refactoring}
M.~Fowler, \emph{Refactoring: improving the design of existing code}.\hskip 1em
  plus 0.5em minus 0.4em\relax Addison-Wesley Professional, 2018.

\bibitem{moreno2014automatic}
J.~Moreno and G.~Robles, ``Automatic detection of bad programming habits in
  scratch: A preliminary study,'' in \emph{2014 IEEE Frontiers in Education
  Conference (FIE) Proceedings}.\hskip 1em plus 0.5em minus 0.4em\relax IEEE,
  2014, pp. 1--4.

\bibitem{santana2024unraveling}
A.~Santana, E.~Figueiredo, and J.~A. Pereira, ``Unraveling the impact of code
  smell agglomerations on code stability,'' in \emph{2024 IEEE International
  Conference on Software Maintenance and Evolution (ICSME)}.\hskip 1em plus
  0.5em minus 0.4em\relax IEEE, 2024, pp. 461--473.

\bibitem{nagappan2006mining}
N.~Nagappan, T.~Ball, and A.~Zeller, ``Mining metrics to predict component
  failures,'' in \emph{Proceedings of the 28th international conference on
  Software engineering}, 2006, pp. 452--461.

\bibitem{ramasamy_workflow_2022}
D.~Ramasamy, C.~Sarasua, A.~Bacchelli, and A.~Bernstein, ``Workflow analysis of
  data science code in public {GitHub} repositories,'' \emph{Empirical Software
  Engineering}, vol.~28, no.~1, p.~7.

\bibitem{huang_robust_2021}
S.~Huang and Q.~Wu, ``Robust pairwise learning with huber loss,'' \emph{Journal
  of Complexity}, vol.~66, p. 101570, Oct. 2021.

\bibitem{cohen2013statistical}
J.~Cohen, \emph{Statistical power analysis for the behavioral sciences}.\hskip
  1em plus 0.5em minus 0.4em\relax routledge, 2013.

\bibitem{tibshirani1993introduction}
R.~J. Tibshirani and B.~Efron, ``An introduction to the bootstrap,''
  \emph{Monographs on statistics and applied probability}, vol.~57, no.~1, pp.
  1--436, 1993.

\bibitem{ACM_SIGSOFT_Empirical_Standards}
\BIBentryALTinterwordspacing
\BIBforeignlanguage{en}{Empirical {Standards} - {Repository} {Mining}}.
  [Online]. Available:
  \url{https://www2.sigsoft.org/EmpiricalStandards/docs/standards}
\BIBentrySTDinterwordspacing

\end{thebibliography}
	
\end{document}